\documentclass[aps,showpacs,twocolumn]{revtex4}
\usepackage{graphicx}%

\usepackage{amsmath}%

\usepackage{amsfonts}%
\usepackage{amssymb}

\usepackage{bm}
\begin{document}
\title{Entanglement and dynamics of spin-chains in periodically-pulsed magnetic fields: accelerator modes }
\author{T. Boness, S. Bose and T.S. Monteiro}\affiliation{Department of Physics and Astronomy, University College London, Gower Street, London WC1E 6BT, United Kingdom}
\date{\today}
%
%
\newcommand{\ket}[1][\psi]{| #1 \rangle}
\newcommand{\bra}[1][\psi]{\langle #1 |}

\begin{abstract}
We study the dynamics of a single excitation in a Heisenberg 
spin-chain subjected to a sequence of periodic pulses from an external,
parabolic, magnetic field. We show that, for experimentally reasonable parameters,
a pair of counter-propagating coherent states are ejected from the centre
of the chain. We find an illuminating 
correspondence with the quantum time evolution
 of the well-known paradigm of quantum chaos, the Quantum Kicked
Rotor (QKR). From this we can analyse the entanglement production and
interpret the ejected coherent states as a manifestation of
so-called `accelerator modes' of a classically chaotic system.

\end{abstract}
\pacs{03.67.Mn,05.45.Mt,05.45.Gg,03.67.Hk} \maketitle

There is considerable interest in the fidelity of quantum state transmission and
 entanglement measures in spin chains because of their relevance
to quantum information applications. In \cite{Bose} state transmission in a
Heisenberg chain was investigated. In \cite{Shi} it was shown that such
a chain, in the presence of an external, static, parabolic magnetic field,
can give perfect transmission of coherent spin states of appropriate width.
Obtaining coherent states of specified widths, represents a technical challenge,
though.

Here, we investigate the dynamics of a Heisenberg spin-chain
subjected to short, {\em time-periodic pulses} from an external parabolic magnetic
field. We find that this provides an effective technique for generating
well-defined coherent states, starting from a single excitation
at the centre of the spin chain. 
The key to the analysis is that we note, for the first time,
a close correspondence between the time-evolution of a Heisenberg chain and that of
the well-known chaotic system the Quantum Kicked Rotor (QKR) \cite{Casati,Fish} in
its quantum resonance regime. 
 Our additional parabolic external field extends the correspondence
between the Heisenberg spin-chain to the non-resonant QKR. 
The non-resonant, chaotic, QKR has been well-investigated experimentally with
cold atoms in optical lattices
\cite{Raizen}. The QKR in the resonant regime has also been investigated experimentally \cite{Summy}.

There is also  much current interest in the interface between quantum chaos 
and quantum information
\cite{Shep2}. In some studies of entanglement measures,
 the quantum chaos is generated by extrinsic disorder
\cite{Cas2}, in others with a clean but chaotic Hamiltonian \cite{Sarkar,Laksh}.
The question of whether chaos aids or hinders entanglement generation relevant
to quantum information applications has not yet yielded a clear answer \cite{Laksh}.
In \cite{Prosen} it was shown that a class of kicked Ising-type chains have
quantum behaviour related to those of one-body `image' systems with a
well-defined classical limit, which can be chaotic or integrable.

But, to our knowledge, the correspondence
 between the dynamics of the QKR,
a leading paradigm of quantum chaos and the Heisenberg chain, a system of such key
interest in quantum information, has not been noted or exploited previously.
We show it means that with the pulsed parabolic field, 
we can employ certain `textbook' \cite{Ott} expressions found 
for the QKR and the Standard Map to describe the entanglement properties.
It means also that we see not only generic
forms of quantum chaotic behaviour in the spin-chain 
like exponential localization (analogous to Anderson localization
 seen in disordered metals) but will also generate phenomena (such
as `accelerator modes') due to additional correlations specific to
a `clean' chaotic system -and to the QKR, in particular. We show that, as we
 can remain in the one-excitation sector with this Hamiltonian,
there are new possibilities for quantum information applications from the
entanglement properties of the accelerator modes.

We consider a time-periodic Hamiltonian of the form:
\begin{eqnarray}
{\bf H}= H_{hc} + \sum_{n=1}^{N} \frac{B_Q}{2}(n-n_0)^2 \sigma_z^n \sum_j \delta(t-jT_0)
\label{eq1}
\end{eqnarray}

where $H_{hc} = -\frac{J}{2} \sum_{n} \sigma^n \cdot \sigma^{n+1} -\sum_{n} B \sigma_z^n$ 
is the  Hamiltonian for the 
Heisenberg chain studied in \cite{Bose}. $T_0$ is the period of the pulses; $B_Q$ is the amplitude 
of the applied parabolic magnetic field; the length of the chain, 
$N \gtrsim 100$ in the present work. The spin-transmission properties for
the time-independent part were investigated in \cite{Bose}: for a non-zero
static field, where $\hbar B \gg kT $ one may restrict the study to 
the single excitation regime
(ie restricted to the basis of states $\ket[ {\bf s} ] $, which have  
a spin-up at a single site $s$ on the chain but all other
spins down (along $-{\hat z}$). 

In \cite{Bose} it was shown that the eigenstates of $H_{hc}$ , $\ket[{\tilde m}]$
are delocalized along the chain ie 
\begin{equation}
    \ket[\tilde m] = a_m \sum_{j=1}^{N} \cos \left[ \frac{\pi}{2N} (m-1)(2j-1)
    \right] \ket[ {\bf j} ]
\label{eq2}
\end{equation}
where $a_m=\sqrt{\frac{2-\delta_{m1}}{N}}$. Using the eigenstates,
an analytical form for the time-evolution
operator $U_{hc}(t,0) = \exp \{ -\frac{i}{\hbar}H_{hc}t \}$ may be obtained
in the single-excitation basis:

\begin{eqnarray}
 \bra[ {\bf r} ] U_{hc}(t) \ket[ {\bf s} ]
 = \sum_{m=1}^{N} a_m^2  e^{ -2iJ t [1- \cos \frac{\pi}{N} (m-1)]}
\nonumber \\
\cos [ \frac{\pi}{2N} (m-1)(2r-1)] 
\cos[\frac{\pi}{2N} (m-1)(2s-1)]\nonumber \\
\label{eq3}
\end{eqnarray}
where we disregard the uninteresting overall phase due to the uniform static field $B$
(or formally set $2BT_0=2 \pi$).
For the periodically-pulsed system described by the Hamiltonian in (\ref{eq1}), the
matrix elements of the one-period time-evolution operator $U(T_0,0)$, 
in the single excitation basis, are:
\begin{equation}
U_{rs}(T_0)= e^{-i \frac{B_Q}{2}(r-n_0)^2} \ \cdot \ \bra[ {\bf r} ] U_{hc}(T_0)
\ket[ {\bf s} ]
\label{eq4}
\end{equation}

\begin{figure}[ht]
\includegraphics*[angle=0,width=3.5in]{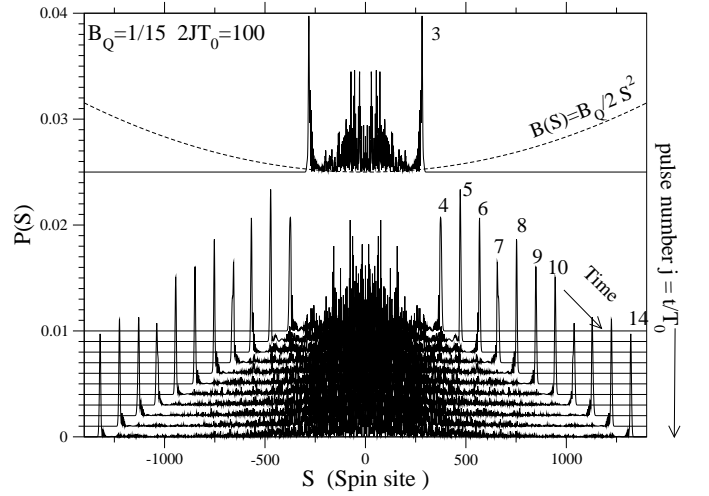}
\caption{Time evolution 
of the state $\psi(t=0)= \ket[ {\bf n_0} ]$ (ie initialized
with a single spin-excitation at the
centre of the chain)- showing the effect of accelerator modes. $P(s)$ represents
the probability of finding the excitation at site $s$.
 The accelerator modes are the `spikes' at the leading edge of the distribution.
They correspond to a counter-propagating pair of coherent states ejected from the centre.  
 We take $B_Q=1/15$ and $2JT_0=100$. The upper line 
is at $t=3T_0$; the lower curves correspond to consecutive periods $jT_0$
with $j=4,5,6...$ as numbered. The dotted line indicates the form of
the parabolic field (scaled by a constant factor)
 which is pulsed on/off every period  at times $t=jT_0$.  
The accelerator modes represent over $25\%$ of the total probability;
they advance an equal distance (shown below to be $ 2\pi/B_Q \simeq 94$ spin sites)
 each period, and after just 3 pulses are well separated from the central, `chaotic'
remnant.}
\label{Fig1}
\end{figure}

In Fig.{\ref{Fig1}}, we show the effect of repeated application of
(\ref{eq3}) and (\ref{eq4})
on the state $\psi(t=0) = \ket[n_0]$ (a state initialized on a site
at, or very near, the centre of the chain. The spin-amplitude spreads out into a
very irregular `chaotic' distribution around the site $n_0$. But most strikingly, we can
see a pair of  counter-propagating spikes, `hopping' around $2\pi/B_Q \simeq 94$ spin-sites each 
 consecutive period. In order to analyse this behaviour, we re-examine the
form of $U^{hc}_{rs}(T_0,0)$.

For large $N$, it is easy to see that (\ref{eq3}) becomes the discretized version of
an integral: as $N \to \infty$, $\bra[r] U^{hc} \ket[s]  \to F_{rs}$ where

\begin{eqnarray}
F_{rs} =\frac{e^{-2iJT_0/\hbar}}{2\pi}  \int_0^{\pi}  
 \left[ \cos(r+s-1) x + \cos(r-s) x \right] \label{eq5} \\ 
e^{\frac{2iJT_0}{\hbar} \cos x}\ dx \nonumber
\end{eqnarray}

The $x=\pi(m-1)/N$ is nominally a position coordinate but in fact represents
motion through the subspace of $\ket[\tilde m]$, the eigenstates of $H_{hc}$.
Eq. (\ref{eq5}) can be compared with the time-evolution operator of one of the
most extensively studied system in quantum chaos, the Quantum Kicked Rotor (QKR) 
\cite{Casati}.  

The QKR corresponds
to the  Hamiltonian $H=\frac{P^2}{2} - K \cos x \sum_n \delta(t-n)$. 
where $K$ is the kick strength and $T=1$ is the kicking period.
The equations of motion for its classical limit produce 
the `Standard Map', the textbook paradigm of classical 
Hamiltonian chaos \cite{Ott}.
The time evolution operator of the QKR, $U^{QKR}$ is
 generally given in a plane wave 
basis \cite{Fish}; ie\\
$\bra[n] U^{QKR} \ket[l] = e^ {-il^2 \hbar/2} \ i^{n-l} J_{n-l}(K/\hbar)$
 where $J$ denotes an ordinary Bessel function. Note that in the so-called
quantum-resonance regime, where $\hbar=2\pi$, (also studied in cold atom
experiments \cite{Summy}) 
we obtain simply  $\bra[n] U^{QKR} \ket[l]  = i^{n-l} J_{n-l}(K/\hbar)$.
In the experiments, $\hbar$ represents an effective value
obtained from the optical lattice parameters and kick period $T$,
  typically of order $\hbar \sim 1$ \cite{Raizen,Jones}.

However,
since parity is conserved, the basis states of the QKR are generally symmetrised, 
$\psi_l^{\pm} =1/\sqrt{2}[\ket[l] \pm \ket[-l] ]$ with $l=0,1,2,...M$, where
$M$ is where we truncate our basis,
and $\ket[l]=1/\sqrt{2\pi} \exp \{ i l x \}$. 

It is evident that, to within an overall phase, the form of the
integral in (\ref{eq5}) is equivalent in form to the `kick' part
of the QKR operator.
\begin{equation}
F_{rs} = \bra[\psi_{r-1/2}^+] e^{i\frac{K}{\hbar} \cos x} \ket[\psi_{s-1/2}] 
\label{eq6}
\end{equation}
if $2JT_0 = K=\beta \hbar$; above, $\ket[r-1/2]$ indicates  plane waves
shifted by half a quantum ie $\ket[r-1/2] =\frac{1}{\sqrt{2\pi}} e^{ i (r-1/2) x}$.
Further,
\begin{equation}
F_{rs} = \bra[\psi_{r-1/2}^+ ] e^{i\frac{K}{\hbar} \cos x} \ket[ \psi_{s-1/2}^+ ] 
 \simeq i^{r-s} J_{r-s}(\beta)
\label{eq7}
\end{equation}
provided we neglect terms which are only significant at the edges of the
basis (ie $r \approx 1$ or $r \approx N$). Note that for a spin-chain on
a ring, these `edge' corrections are  entirely absent and the time evolution
of a Heisenberg chain ring is entirely equivalent to the kick part of the QKR.
The addition of the parabolic-field `kick' completes the analogy with the full QKR: 
the fact that the parabolic field is now the $\delta-kick$ term, while for the QKR
the kinetic energy provides the time-independent term, represents only a simple re-scaling of
the parameters in $U(T_0,0)$. 

\begin{figure}[ht]
\includegraphics*[angle=0,width=3.in]{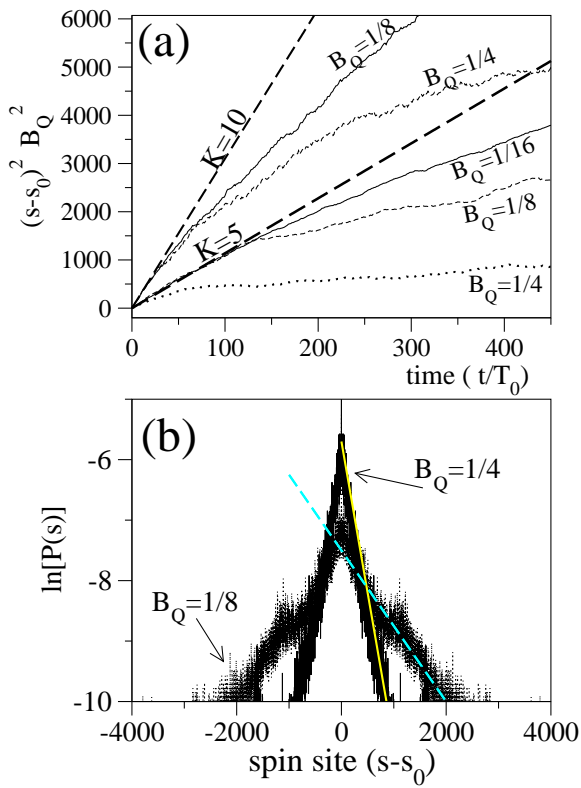}
\caption{{\bf (a) }Shows that, the rate of spreading of a spin excitation initially
 at an arbitrary site $\ket[s_0]$ in the chain, at short times, is determined solely
 by the `classical stochasticity parameter' $K_s =2JT_0 B_Q$. At longer times, the 
spreading saturates. The straight line indicates the behaviour expected of 
 the classical chaotic kicked rotor. The smaller
the effective value of Planck's constant, $B_Q$, the longer the spin
spreads linearly at the `classical' rate. Note that in the accelerator mode regime
however, the increase would be quadratic in time, not linear.
{\bf (b) } Shows that at sufficiently long times the spin probability distribution saturates
into an exponentially localized form. This is the analogue of Anderson Localization in
a disordered metal.  $P(s) \sim \exp \{-{2|s-s_0|/L} \}$ so  $ln(P(s)) \sim 2/L$
takes the characteristic `triangular form'. The straight lines correspond to $L= \frac{K_s^2}{4B_Q^2}$
for $K_s=10$ and agree well with the numerically calculated distributions.}
\label{Fig2}
\end{figure}

Note that all our numerics employ (\ref{eq3}) and not the QKR form in
(\ref{eq7}) above.
Nevertheless, we found the spin-chain dynamics are sufficiently analogous that we can simply
make the substitutions: (a) $B_Q= \hbar$ is an effective value of Planck's constant.
(b) $ 2JT_0 B_Q=K$, where $K$ is the stochasticity parameter which
fully determines the classical dynamics of the kicked rotor; eg
 $K \gtrsim ~4.5 $ indicates fully chaotic dynamics. 
We can then directly apply the well-known QKR results to the
spin dynamics:

1)\emph{Short-time diffusion of spin-excitation}:  For $K \gtrsim 4$ the QKR
initially displays diffusion of momentum, $\langle p^2 \rangle \approx D(K) \ t$,
which follows the classical behaviour: the
 form  $D(K) \approx K^2/2 \{ 1-2J_2(K)+2{J_2(K)}^2 \}$ \cite{Rechester}
is obtained from a study of the classical diffusion for the Standard Map. The first term in
the expression for $D$ represents uncorrelated diffusion (equivalent to a random
walk); the $J_2(K)$ terms result from short-time  correlations present in the
classical kicked rotor. 
 We obtain analogous behaviour in
our pulsed spin chain; substituting $2JT_0B_Q = K_s$, and starting out with a spin
initialised on a site $s=s_0$ {\em anywhere} on the chain, we find $\langle
(s-s_0)^2 B_Q^2 \rangle \approx D(K_s) t$. The  diffusion of spin excitation obtained 
from the time-evolution operator in Eq.\ref{eq3} is compared, in
Fig.\ref{Fig2}(a),  with the form of the classical diffusion rate of the QKR. 
  For $K_s = 10$, the correction due to short-range correlations
is large, but for $K_s = 5$, we find $D \approx K_s^2/2$  since $J_2(5)
\approx 0$.  After the so-called `break-time', $ t^* \sim (K_s/B_Q)^2$ the quantum
diffusion no-longer follows the classical behaviour and the spin-excitation
stops spreading onto neighbouring sites.

2) {\em Dynamical Localization}: The Floquet states (eigenstates of $U^{QKR}(T,0)$ ) of the QKR are known to 
be delocalized in the resonant regime; hence, as investigated in \cite{Bose},
a spin-excitation can spread (imperfectly) from one end of the chain to the other.
 In the non-resonant case, however, the corresponding Floquet states exhibit Dynamical
Localization, a phenomenon analogous to Anderson Localization in a disordered
metal \cite{Fish}. Hence, for the spin-chain in pulsed parabolic field, a
single excitation $anywhere$ in the chain will spread up to a maximum
and will `freeze' (localize) at the break-time,
$t^*$. For $t > t^*$,  
\begin{equation}
    P_{loc}(s) \sim \frac{2}{L} \exp \{ -2|s-s_0|/L \}.
    \label{eqPLOC}
\end{equation}
The quantum localization length $L \simeq (2JT_0)^2/4$. Fig. \ref{Fig2}(b)
demonstrates the typical probability distribution of a dynamically localized state,
obtained by evolving a state initially in an arbitrary site.

3){\em Entanglement}: Our knowledge of the time evolution and spreading of
 the spin distribution enables us to estimate some entanglement measures;
in particular, the $Q$-measure \cite{Meyer} and the Concurrence \cite{Woott}.
The former is a measure of the global entanglement for a pure (multipartite)
state:  
\begin{equation*}
    Q(\ket)= \frac{4}{N}\left( 1-  \sum_{k=1}^N  |\alpha_k|^4 \right)
	    \simeq \frac{4}{N} \left(1- \frac{1}{L} \right)
\end{equation*}
where the $\alpha_k$ are the amplitudes when the state is projected on to the
basis of spin sites, $\{ \ket[k] \}$.  We note the relation of this formula to the
well known Inverse Participation Ratio, $R=1/\sum |a_k|^4$; for a localised
system $R \simeq 2L$.

We assign a measure to the bipartite entanglement of two separated sites
$i$ and $j$ with the Concurrence.  Given a pure state in the single excitation
basis $C_{i,j} = 4|\alpha_i| |\alpha_j|$ \cite{Fitz}. With the exponentially
localized form, (\ref{eqPLOC}):
\begin{equation*}
    C_{i,j} \sim \frac{8}{L} \exp \left\{-\frac{2}{L}\left(|i - s_0| + |j - s_0| \right)
\right\} 
\end{equation*}
if the separation between two sites $d_{i,j}=|i - s_0| + |j - s_0| > 0$,
 the maximum of the concurrence occurs at $L=2d$,
with the value $\frac{4}{d} e^{-1}$.

4){\em accelerator modes}: For values of $K \approx 2\pi$, transporting islands
of stability re-appear in the classical phase space. Their phase-space area is 
$\sim 1/10$ and hence can support quantum states if $ \hbar \lesssim 1/10$.
Their effect has been experimentally observed for cold cesium atoms in 
pulsed standing waves of light \cite{Raizacc}. Because of the relatively 
large effective values of $\hbar \gtrsim 1$  in the atomic experiments, their
effect there was diluted by the chaotic `sea' of trajectories;
 nevertheless, the accelerator modes (and Levy
flights due to their environs) manifested themselves as an enhancement of the diffusion
for $K \simeq 2\pi$.

In the spin-chains there is no evident bar to a low $\hbar=B_Q$: it simply requires
a weaker magnetic field. Accelerator modes, in fact, are stable for a broad 
parameter range $K = \alpha 2\pi$ where $ \alpha=1.03 - 1.10$ so there is
no need for very precise fine tuning of the parameter $2J T_0$.
 The associated islands of stability
span a width (of spin sites) of order $\Delta s \sim \frac{1}{10} 2\pi/\hbar$,
(about $10$ sites for Fig. \ref{Fig1}, so there is no need
to prepare the initial state {\em exactly} at the centre of the chain either, in order
for the initial state to overlap strongly with the accelerator islands.

We find useful coherent states are obtained provided $B_Q \lesssim 1/5$. The width of
these states is simply determined by the effective $\hbar$ : we verified numerically
that their form is given by  $\psi(s, t=jT_0) \simeq  A_j\exp \{ -B_Q(s-s_j)^2
\}$ where
$s_j=2\pi j/\hbar$ and $j$ is the pulse number. We find that, for
$j \simeq 1$ the coherent state pair represents about $30 \%$ of the total probability. For finite
$\hbar$ we expect the amplitude to gradually `tunnel' out of the accelerator island:
for $\hbar =1/10$ we estimated the decay numerically $|A_j|^2 \simeq |A_{j=1}|^2
\exp \{ -j/24\}$, so even after
 30 pulses there is a substantial amplitude. For $\hbar =1/15$ (shown in Fig. \ref{Fig1}) the
island can in fact support more than one eigenstate, leading to a slight oscillation
in its amplitude.  It only requires $2-4$ pulses to cleanly separate the travelling
states from the chaotic remnant, though, so even $\hbar \simeq 1/5$ would give coherent states
with probability $\geq 20\%$. These could then be taken into a static, parabolic field region
and transmitted onwards with perfect fidelity \cite{Shi}.

For an actual realization we might suggest $B_Q \sim 1/10$; in an experiment, $B_Q = {\tilde B_Q} \delta t$
where $\delta t \ll T_0$ is the pulse duration. If our maximum magnetic field $B_{max}$ is of order
0.1 Tesla, 
$B_{max} \sim (N/2)^2 {\tilde B_Q} \sim 10^{-6}\ au$ , for $N \sim 100 -1000$, 
implies  ${\tilde B_Q} \sim 10^{-10}\ au$
so the pulse duration $\delta t \sim 10^{9}\ au \sim 25 \ ns$. As we have made a split-operator
approximation in Eq. (\ref{eq4}), we require $2J \delta t $ is a small phase, whereas
 $2J T_0$ is significant. This constrains $2J \sim 10^{-9} \sim 1-10 \ MHz$, so we
are in a weak spin-coupling regime, for this choice of parameters.

We now propose an application of the accelerator modes in
context of quantum communication. It has been noted that encoding
quantum information in Gaussian wave-packets of excitation (where
the presence and absence of the wave-packet depicts logical
$|1\rangle$ and $|0\rangle$ states of a qubit) is a useful way of
transmitting it down a Heisenberg coupled chain of spins
\cite{osborne,Shi}. However, precisely {\em how} to create a
superposition of the presence and absence of a Gaussian wave-packet
remained unclear. The work presented here suggests, in fact,
 that after 3-4 kicks one creates a superposition of two
Gaussian wave-packets traveling in opposite directions and a
exponentially localized state in the middle of a Heisenberg spin
chain. If one measures the exponentially localized part and does not
find the spin excitation there (which can have nearly $30\%$
probability of occurrence for appropriately chosen parameters), then
the spin chain is projected on to a superposition of oppositely
traveling Gaussian wave-packets of excitation. If one denotes the
left and right traveling wave-packets as $|G_L\rangle$ and
$|G_R\rangle$ (which are each a Gaussian distribution of spin up),
and $|0_L\rangle$ and $|0_R\rangle$ as the absence such a
wave-packet (which are each an all spin down state at the sites
where $|G_L\rangle$ and $|G_R\rangle$ would otherwise be) then the
maximally entangled state $|G_L\rangle|0_R\rangle+|0_L\rangle|G_R\rangle$ is
effectively created. This state can now be distributed among well separated
parties by switching on a constant parabolic field (instead of the
kicks) in which a Gaussian wave-packet can travel for a significant
distance without distortion \cite{Shi}. \\
 In conclusion: we have demonstrated that there is a close correspondence between a 
spin-chain in a pulsed quadratic field and the well-known chaotic kicked rotor;
 we find potential applications in quantum information processing. \\
{\em Acknowledgements}: Tom Boness acknowledges a PhD studentship from the EPSRC.

\end{document}